\documentclass{abpaper}
\usepackage{subfigure}   

\usepackage{xspace,amsmath,booktabs,graphicx}
\newcommand{\pT}{\ensuremath{p_\perp}\xspace}
\newcommand{\kT}{\ensuremath{k_\perp}\xspace}

\newcommand{\sqrtS}{\ensuremath{\sqrt{s}}\xspace}
\newcommand{\alphaS}{\ensuremath{\alpha_\text{S}}\xspace}

\newcommand{\chisq}{\ensuremath{\chi^2}\xspace}

\renewcommand{\vec}[1]{\ensuremath{\mathbf{#1}}\xspace}
\newcommand{\p}{\vec{p}\xspace}

\newcommand{\ptilde}{\ensuremath{\tilde{p}\xspace}}

\newcommand{\professor}{\textsc{Professor}\xspace}

\newcommand{\pythia}{\textsc{Pythia}\xspace}
\newcommand{\pythiasix}{\textsc{Pythia\,6}\xspace}
\newcommand{\pythiaeight}{\textsc{Pythia\,8}\xspace}
\newcommand{\herwig}{\textsc{Herwig}\xspace}
\newcommand{\herwigpp}{\textsc{Herwig\raisebox{0.2ex}{++}}\xspace}
\newcommand{\herwigsix}{\textsc{Herwig\,6}\xspace}

\newcommand{\jimmy}{\textsc{Jimmy}\xspace}
\newcommand{\sherpa}{\textsc{Sherpa}\xspace}
\newcommand{\powheg}{\textsc{Powheg}\xspace}

\newcommand{\tevatron}{\textsc{Tevatron}\xspace}

\newcommand{\gof}{\textsc{GoF}\xspace}

\newcommand{\MeV}{\text{Me\kern -0.15ex V}\xspace}
\newcommand{\GeV}{\text{Ge\kern -0.15ex V}\xspace}
\newcommand{\TeV}{\text{Te\kern -0.15ex V}\xspace}

\newcommand{\fref}[1]{Figure~\ref{#1}}
\newcommand{\tref}[1]{Table~\ref{#1}}
\usepackage{caption}
\makeatletter
\DeclareOldFontCommand{\rm}{\normalfont\rmfamily}{\mathrm}
\DeclareOldFontCommand{\sf}{\normalfont\sffamily}{\mathsf}
\DeclareOldFontCommand{\tt}{\normalfont\ttfamily}{\mathtt}
\DeclareOldFontCommand{\bf}{\normalfont\bfseries}{\mathbf}
\DeclareOldFontCommand{\it}{\normalfont\itshape}{\mathit}
\DeclareOldFontCommand{\sl}{\normalfont\slshape}{\@nomath\sl}
\DeclareOldFontCommand{\sc}{\normalfont\scshape}{\@nomath\sc}
\makeatother

\usepackage{afterpage}

\title{Tuning of MC generator MPI models} 
\author{
  \textbf{\sffamily Andy Buckley}\\[-0.5ex] \textsmaller{\rmfamily\itshape School of Physics \& Astronomy, Glasgow University, Glasgow, UK}\\[1ex]
  \textbf{\sffamily Holger~Schulz}\\[-0.5ex] \textsmaller{\rmfamily\itshape Department of Physics, University of Cincinnati, Cincinnati, OH 45219, USA}}

\begin{document}



\begin{abstract}
  \textit{Contributed chapter to ``Multiple Parton Interactions at the LHC'', World Scientific}\\[0.8ex]
  MC models of multiple partonic scattering inevitably introduce many free
  parameters, either fundamental to the models or from their integration with MC
  treatments of primary-scattering evolution. This non-perturbative and
  non-factorisable physics in particular cannot currently be constrained from
  theoretical principles, and hence parameter optimisation against experimental
  data is required. This process is commonly referred to as MC tuning. 
  We summarise the principles, problems and history of MC tuning, and the
  still-evolving modern approach to both model optimisation and estimation of
  modelling uncertainties.
\end{abstract}


\section{Introduction}
\label{sec:tuning:intro}

It is an unfortunate fact of life that the modelling approaches to multiple
partonic scattering 
implemented in Monte Carlo event generator programs are not fully
unambiguous. Not only are ans\"atze required for computation of the hadronic
matter overlap and low-\pT regularisation of the partonic scattering
cross-section, but also the secondary scattering must coherently connect to the
other aspects of event modelling in general-purpose MC codes.

For example, MPI scattering must interface somehow with QCD evolution down to
soft momentum-transfer scales, as described by parton showers and matrix element
corrections, and to the colour connections between the partonic scattering
system and the beam remnants. Phenomenological hadronisation models must also be
modified to accommodate MPI as a source of partons, most notably via somewhat
\textit{ad hoc} ``colour reconnection'' or ``colour disruption'' mechanisms.

The result of this complexity is that MPI models not only contain degrees of
freedom intrinsic to their own formulation, but also require extensions to the
generator components concerned with the primary partonic scatter. As much of the
physics involved is non-perturbative --- and that which is not is only defined
up to leading-order or leading-logarithmic accuracy --- it is typical for more
than ten model parameters to influence observables of interest, with little or
no \emph{a priori} prediction of their values. These parameters must somehow be
``tuned'' to describe MPI-sensitive observables in experimental data.

In this chapter we describe the dominant modern approach to MC generator
parameter optimisation, from the technical machinery to the parameters and
observables, and the methodology applied to both achieve convergence and avoid
overfitting. As is always true for parameter estimation in physics, the
resulting uncertainty is as important as the central value and hence we review
the statistical methodology applied to estimate model uncertainties through
tuning. Finally, we survey the road ahead for MC tuning and improved constraints
on MPI modelling.

\section{Tuning methodology}
\label{sec:tuning:method}

From the outset we should be clear that tuning is not desirable. While currently
a necessary part of the landscape of MPI modelling (and other hadron-collider
event features), in an ideal world our models would have sufficiently few
ambiguities that tuning will become unnecessary. We can dream!  But for now,
modelling flexibility is necessary to achieve the degree of data description
required by experiment --- at the significant cost of exchanging parameter
fitting and uncertainty estimation for predictivity.



In this pragmatic compromise, it is preferable not to simply throw all possible
parameters and data into a massive fit. Rather, well-motivated modelling
components --- typically those involving QCD at perturbative scales --- are
trusted to be predictive from first principles, while phenomenological models
which are only unconstrained in the asymptotic limits of QCD are ripe for
fitting. We use the word ``tuning'' to refer only to fits of the latter
parameter type, and as far as possible avoid fitting true theory uncertainties
such as scale choices in the perturbation expansion.


In the simplest case, tuning consists of finding the value of a single model
parameter --- say, the \pT scale used to regularise the divergent secondary
scattering cross-section --- which gives the best agreement with a single data
bin, e.g.~a total cross-section for double-partonic scattering. For this, little
technical machinery is required: a set of MC generator runs with different
values of the parameter (either over the whole natural range of the model, or
focused on a ``known-good'' region) are compared to the data,
and the best-performing model point is chosen as the ``tune''. Perhaps a
couple of iterations will be required. This is ``manual tuning.''

Extension of this scheme to include more data is simple but not entirely
trivial. For example, a multi-bin observable such as an ``underlying event''
characterisation of mean particle or energy flow away from the hard primary
scattering products, simply requires that the goodness of fit be computed via an
aggregation of the fit quality across the many bins. Computing the fit quality
naturally introduces several questions:
\begin{enumerate}
\item Which regions of the observable are most important to describe?
\item Are there bins which the model fundamentally can/should not describe?
\item Are these bins independent of each other, or correlated somehow?
\end{enumerate}
Unfortunately we do not have general answers to the first two of these points,
which immediately makes a robust statistical foundation for tuning
problematic. The third in principle can be solved by publication of
bin-correlation data, but this has both been rare in MPI-sensitive measurements
until now, and is arguably rendered moot by the first two issues. We shall
return to this theme later.

\begin{figure}[t]
  \centering
  \includegraphics[width=0.37\textwidth]{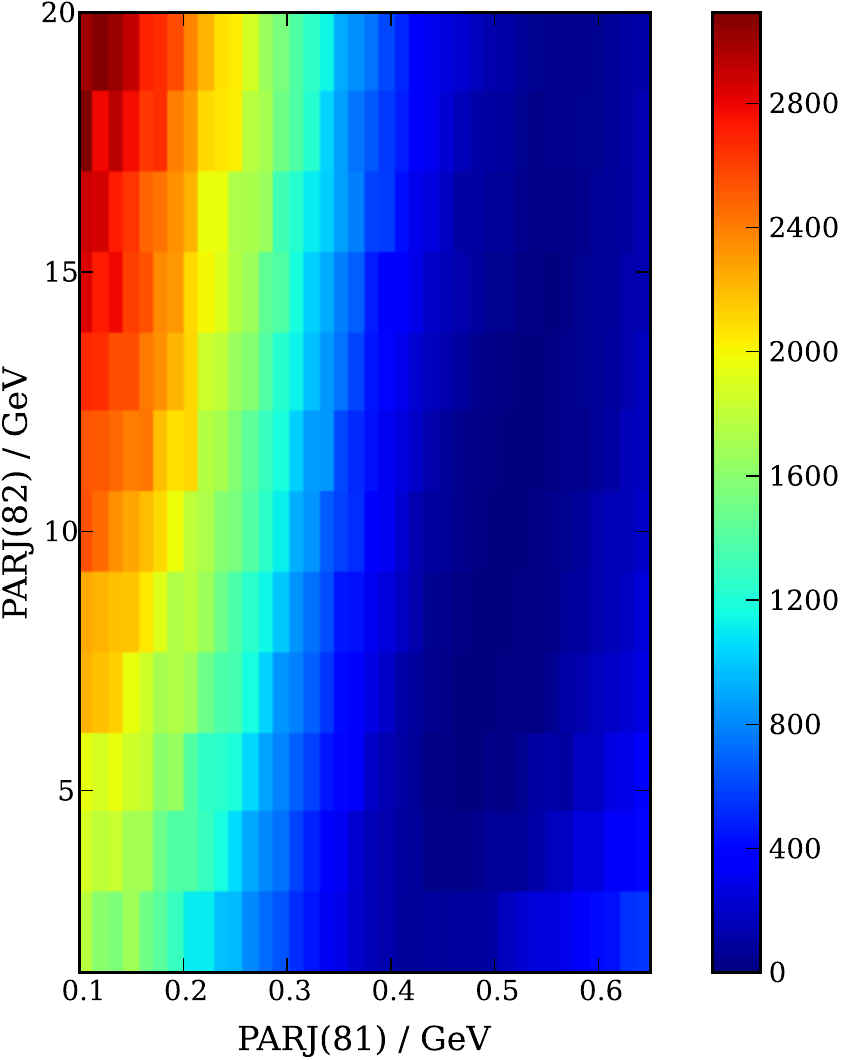}\qquad
  \raisebox{1ex}{\includegraphics[width=0.55\textwidth]{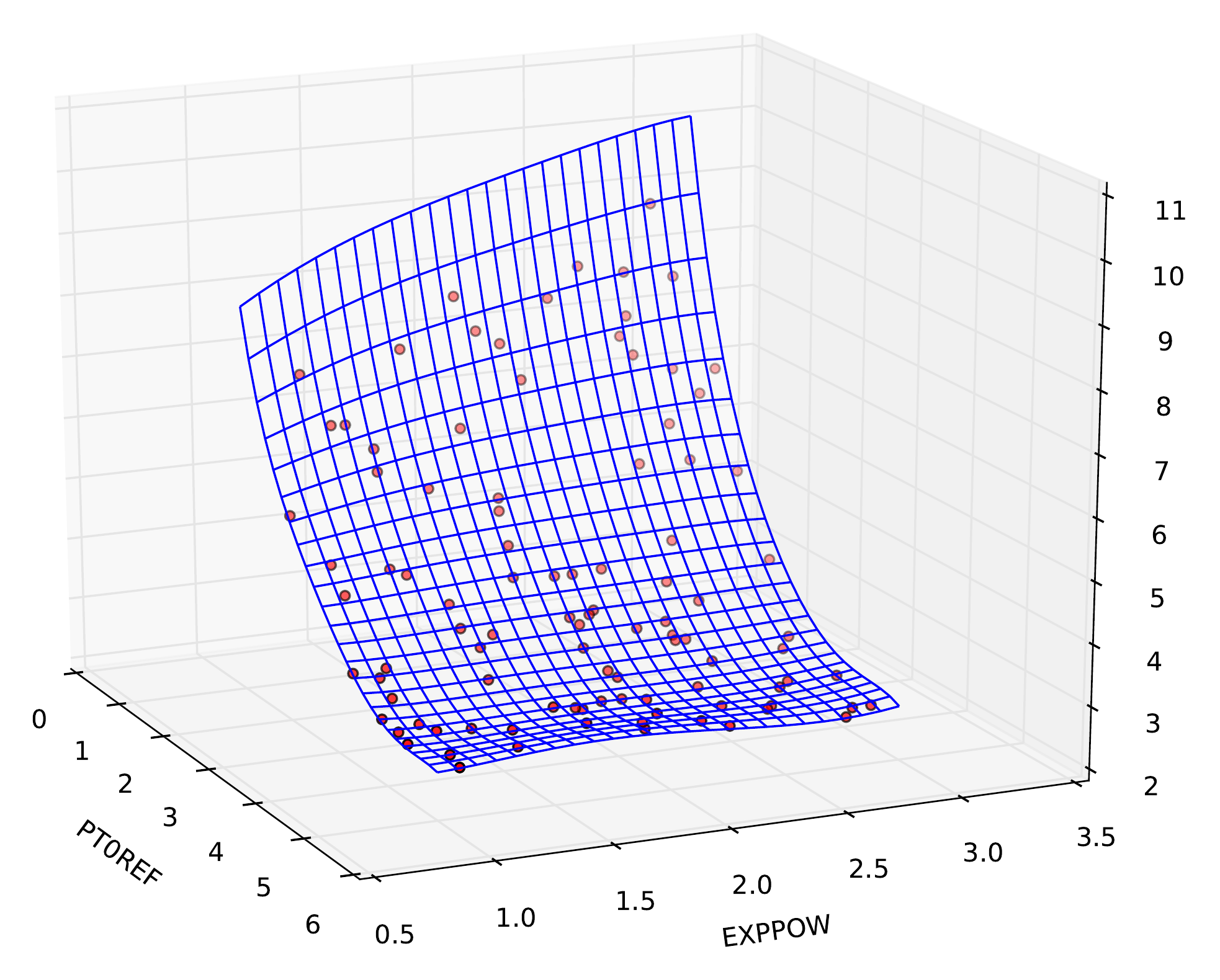}}
  \caption{Goodness-of-fit characterisations in 2-parameter tunes. Left, LEP
    event thrust in \pythiasix parton shower; Right, LHC minimum-bias in
    \pythiaeight MPI model (with \professor $\chi^2$ function fit to the sampled
    points).}
  \label{fig:tuning:2dchi2}
\end{figure}

Now a more technically troublesome extension: more \emph{parameters}. In a
two-dimensional parameter space, such as including a scaling parameter 
for the hadronic matter distribution in addition to the \pT regulariser --- a
by-eye approach is still possible: as before, make MC runs for several
combinations of the two parameters' values and compare to data. In addition,
computing a goodness-of-fit measure for all points in a 2D grid can provide a
useful visualisation of the physics dependence, as illustrated in
\fref{fig:tuning:2dchi2}. But the increased computational cost is clear: if $N$
trial points were required for a single parameter, $\mathcal{O}(N^2)$ will be
required for two parameters, and this exponential scaling continues into the
less visualisable spaces of 3 and more parameters. The severity of this problem
is emphasised by the expense of a typical parameter-point evaluation:
the required statistical precision means that $\mathcal{O}(\text{100k--1M})$
events are needed per colliding-beam configuration and per primary-scattering
process type, hence the total time for a parameter point may be counted in
\emph{CPU-days}. A comprehensive grid-scan of the $\mathcal{O}(\text{10--20})$
parameters needed in a full generator is clearly unfeasible.

One possible approach is physical intuition, and this is undoubtedly
helpful. Fortunately a 20-dimensional appreciation of model behaviour is not
necessary, since many modelling components approximately factorise and can be
reasoned about in isolation. But even so, once more than a few interconnected
parameters are involved, intuition is not enough to break fit-quality
degeneracies, and it is impossible to be sure that a final intuitive tune is
truly optimal. However, despite the availability of more technical machinery,
the wholly intuitive approach is far from irrelevant~\cite{Skands:2014pea}.

The alternative is to somehow make do with an incomplete sampling of the
parameter space, and to use computational machinery to guide the fit. Done
na\"ively, this runs into problems of its own: a random sampling of a
large-dimensional space gives little confidence that the best-seen parameter
point is anywhere near to the global optimum. Attempting to
systematically improve on the points visited so far
would provide a better sampling of the space, but at the unsustainable cost of
abandoning the implicitly parallelisable approach of independent MC runs.
Even very sophisticated attempts to serially sample MPI model parameter spaces
have proven intractable due to the high computational cost of the MC
``function''~\cite{Kama:2010bva}. Instead, the method which has become most
widespread, via the \professor toolkit~\cite{Buckley:2009bj}, is a hybrid of
parallel sampling and serial optimisation, via \emph{parametrisation} of the MC
generator response to parameter variations. It is this approach to which we will
dedicate most space in this summary. First, however, it is important that we
acquaint ourselves with the typical parameters and observables that will be
encountered while constructing an MC generator tune.

\subsection{Experimental observables and model parameters}
\label{sec:tuning:obsparams}

While our interest here is naturally biased toward the MPI-specific
aspects of generator modelling, it is worthwhile to also discuss how the rest of
the generator system is involved in tuning. In rough reverse order from final to
initial state, the main components of an MC generator are hadronisation,
fragmentation, final-state QCD radiation, initial-state QCD radiation, and
multiple partonic scattering. Taken together, these components readily comprise
a parameter space with in excess of 30 dimensions. Even with computational
tricks in the generator runs, sampling such a space requires extraordinary CPU
resources. But there are usually significant factorisations between these major
modelling steps, which can be exploited to make the tuning more tractable.

Typically we start by identifying these factorised blocks working backward from
the final state, assuming that decays, hadronisation (not including colour
reconnection\footnote{Colour reconnection (CR), i.e. dynamic reconfiguration of
  final-parton colour string/cluster topologies, is more a hadronisation effect
  than an MPI one although it is often discussed as an aspect of secondary
  scattering. In principle it can therefore affect final-state observables,
  although there is as yet no strong evidence for this. More concerning is that
  soft-QCD tunes of CR may have inappropriate effects in high-\pT
  hard-scattering processes e.g. $t\bar{t}$ production.}), and the final-state
parton shower should be sufficiently independent of initial-state QCD effects
that they can be tuned to $e^+e^-$ data from LEP and SLD only. Using hadron
collider data (at least in the first iteration) could bias this tuning via poor
initial-state modelling, so it is typically left out, saving CPU time and
complexity. Since hadronisation models can account for the majority of MC
generator parameters, it is not unusual to split this final-state tuning stage
into several rounds, e.g. first concentrating on using identified hadron
rates~\cite{Olive:2016xmw} to fix hadronisation parameters such as strangeness
suppression, then using identified particle energy spectra, event shapes, and
jet rates~\cite{Abreu:1996na,Barate:1996fi,Pfeifenschneider:1999rz} to constrain
the final-state parton shower and fragmentation functions. As the parton shower
is built on perturbative physics, it has few degrees of freedom: typically just
the emission cutoff scale and perhaps the definition of \alphaS evolution.
Treated in this combined way, the first \professor tunes of \pythiasix achieved
degrees of data description for LEP event shapes, jet rates, and
$b$-fragmentation that had previously eluded manual tunes.

Having addressed the tuning of final-state modelling, we now confront the
combination of the initial-state parton shower, the MPI mechanism, and the
colour-reconnection mechanism: all key to description of particle production and
energy flow observables at hadron colliders. The most important observables
depend on the bias of one's physics interest, but in the general-purpose tunes
which have received most attention, the emphasis in soft QCD has been on
kinematics rather than flavour content. 
This due to experimental priorities: for most purposes at the LHC,
secondary-scattering is a background to measurement of hard signal-process
scattering, and hence the key requirement of a general-purpose MPI tune is to
accurately model its contribution to charged track multiplicities and energy
flows. Since these background contributions are divided into inclusive soft-QCD
scattering in additional $pp$ collisions (known as ``pile-up''), and additional
partonic scattering in signal events (the ``underlying event''), the dominant
observables in such tunes are those measured in inclusive ``minimum bias'' event
selection, and those specific to the underlying event.

Minimum-bias physics measurements at hadron colliders have been dominated by
charged-track observables, partially because the low-luminosity early phases of
each data-taking run are crucial for calibration of detector tracking
systems. The main data from these measurements, often grouped as ``min-bias
observables'', are charged-particle multiplicities and \pT spectra within
fiducial acceptance cuts (most obviously the tracker coverage in pseudorapidity,
$\eta$), and the correlation of the average charged-particle \pT with the
event's fiducial track multiplicity. Broadly speaking, the distribution of
charged-particle multiplicity $N_\mathrm{ch}$ with $\eta$ is the canonical
distribution used to indicate the inclusive amount of minimum bias particle
production, particularly in the flat central region $|\eta| < 1$. This is
governed by a correlated combination of MPI $\pT^0$ regularisation parameter,
the amount of hadronic matter overlap for large impact parameter (since
$b \sim 1/Q$ and the typical scale $Q$ of inclusive minimum-bias interaction is
low), and perhaps scaling of the MPI partonic process via 
freedom in parametrisation of the MPI \alphaS. The correlation of
$\langle \pT \rangle$ with $N_\mathrm{ch}$ became an important tuning observable
when it was noted that models without colour-reconnection were unable to
describe it, predicting too soft a particle production spectrum in
higher-multiplicity scattering events. 
Colour disruption mechanisms were added to the hadronisation models of \pythia and
\herwig to address this
, naturally introducing
extra degrees of freedom for tuning.

In the LHC era, additional fiducial cuts were introduced to minimum-bias data
analyses as variations of ``analysis phase space'' to modify the sensitivity to
different aspects of inclusive scattering physics. These include requirements on
charged-particle fiducial multiplicity (e.g. from 1 to $\sim \! 20$) and on minimum
charged-particle~\pT (e.g. from 100~\MeV to $\sim \!\! 10$~\GeV). The result is that
a very large number of many-bin observables, generally with small uncertainties,
is available from the Tevatron to the LHC, and from 300~\GeV to 13~\TeV beam
energies~\cite{Abelev:2012sea,Abe:1989td,Acosta:2001rm,Aaltonen:2009ne,Aad:2010rd,Aad:2010ac,Aad:2016mok,Aad:2016xww,Aaboud:2016itf,Khachatryan:2010nk}.
In addition, ATLAS and CMS have published calorimetric measurements of energy
flow as a function of $\eta$, including a minimum-bias trigger selection. These
provide a counterpart to the track-specific measurements, including both a
central overlap with tracking detector acceptance and extension to
high-$|\eta|$, crucial for forward/diffractive and beam-connection physics. This
availability of multiple independent measurements of each observable is
important to avoid overfitting of a single measurement, and the different phase
spaces enable, for example, degeneracy breaking between non-diffractive MPI and
diffractive physics contributions to particle production.

Underlying event (UE) analyses are a specialisation of the minimum bias
observables to events where a more exclusive trigger is required, i.e. a
genuinely high-scale scattering process such as hard jet or $Z$ boson
production. Typically the motivation of UE measurements is to specifically study
the connection between this hard process and the associated secondary
scattering, as a test of the eikonal MPI model and of the interaction between it
and the perturbative QCD dressing of the primary scatter.

Since a hard primary partonic process will dominate the particle- and
energy-flow characteristics of each event, UE analyses specifically analyse
event regions expected to contain minimal hard process contamination --- for
example, regions azimuthally transverse to the axis of a balanced dijet
event~\cite{Aad:2014hia,Chatrchyan:2011id,Khachatryan:2015jza}, transverse to or
in the direction of a hard leptonic $Z$~\cite{Aaltonen:2010rm,Aad:2014jgf}, or
with identified jet activity ``cut out'' from anywhere in the event
$\eta$--$\phi$ phase space~\cite{Acosta:2004wqa}.  The transverse regions are
often further specialised to discriminate between the more and less active sides
on a per-event basis, to provide additional resolution between MPI and parton
shower activity. 

The canonical UE observable is the evolution of the mean value of a minimum-bias
event property like charged-particle multiplicity or \pT sum within a sensitive
phase-space region, as a function of the scale (usually a \pT) of the hard
scattering process. This produces an extremely informative curve showing the
smooth evolution of mean event properties from minimum-bias at low event scales
(interpreted as peripheral hadronic collisions), up to very hard primary
scatterings 
as $b \to 0$. Underlying event physics hence probes the same MPI mechanisms as
minimum-bias (particularly high-activity MB phases spaces), but with a clear
connection to the pedestal effect, which maps the matter overlap profile in
detail, and an increased emphasis on the role of initial-state QCD radiation
from the high-scale primary process.

\begin{figure}[t]
  \centering
  \subfigure[][]{\label{fig:tuning:pdflumipt0}\includegraphics[width=0.48\textwidth]{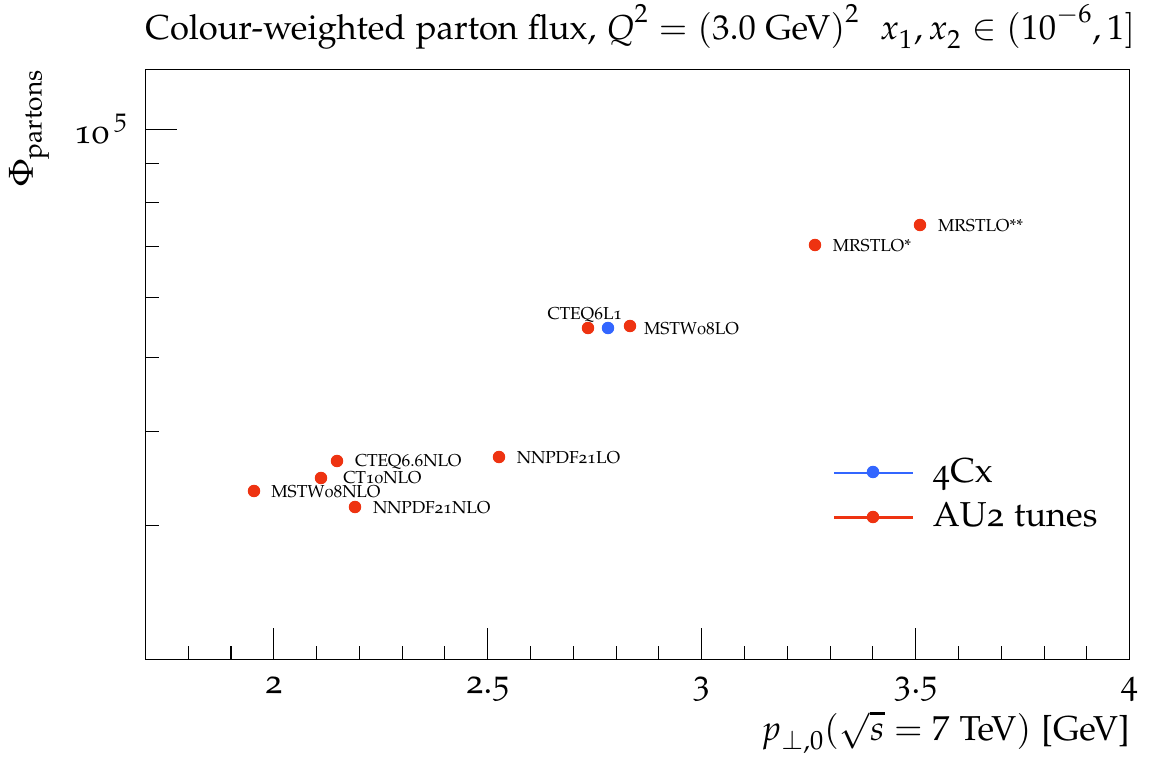}}
  \subfigure[][]{\label{fig:tuning:pt0sqrts}\includegraphics[width=0.48\textwidth]{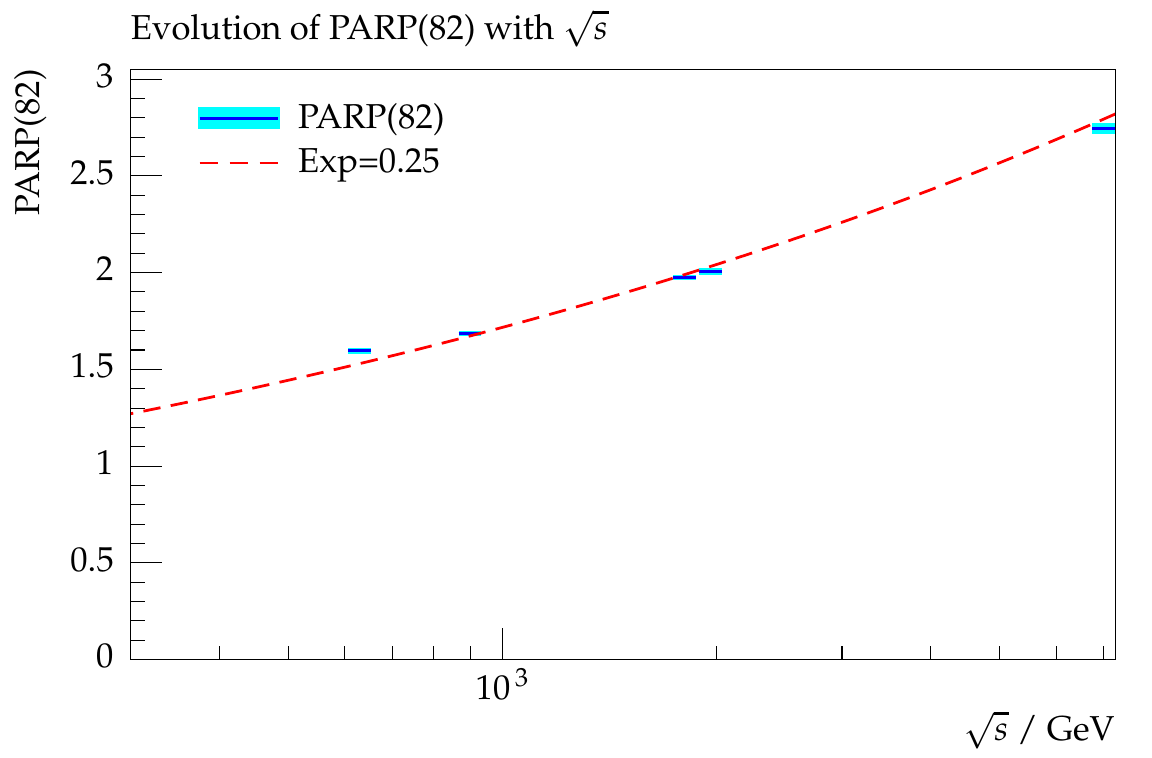}}
  \caption{\subref{fig:tuning:pdflumipt0}
    Tuned \pythiaeight MPI cross-section regulariser, $\pT^0$, as a function of
    low-$x$ gluon luminosity from various PDFs. %
    The correlation between higher gluon flux and higher $\pT^0$ (meaning more
    screening of that flux) is clear. \subref{fig:tuning:pt0sqrts} Empirical
    $\pT^0$ scaling via independent tunes to hadron collider data at different
    \sqrtS.}
  \label{fig:tuning:pt0vspdflumisqrts}
\end{figure}

Parton density functions play an important role in hadronic initial-state
tuning, particularly MPI\footnote{Initial-state parton showers are largely
  insensitive to PDF detail, at least for a given \alphaS and perturbative order
  in QCD~\cite{Buckley:2016caq}.} since the partonic cross-section for multiple
scattering at low-$Q$ is strongly driven by the diverging low-$x$ gluon PDF,
which varies a great deal between different PDF fits. Additionally, PDF
differences produce variations in the rapidity distribution of MPI partonic
scattering. These effects, influencing the multiplicity of MPI scattering and
$d\eta/dN_\mathrm{ch}$ distribution shapes respectively, are responded to in
tuning by correlated shifts in the $\pT^0$ screening factor, matter
distribution/overlap parametrisation, and MPI \alphaS.  The \pythia MPI model in
particular is rather over-parametrised: the MPI rate can be more-or-less
directly modified via $\pT^0$, \alphaS, \emph{and} a partonic scattering
scale-factor. Na\"ively throwing all these parameters into a fit will likely
produce degenerate or overfitted tunes: it is best to instead use a subset of at
most 3 parameters, e.g. \alphaS or scale-factor (or neither) but certainly not
both in a single tune. The relationship between gluon luminosity (the integral
of a PDF over MPI partonic scattering scales) and tuned $\pT^0$ may be seen in
\fref{fig:tuning:pdflumipt0}: a more divergent low-$x$ PDF with a higher
gluon luminosity is strongly correlated with a higher $\pT^0$ tune value and
hence more screening of the divergence. Tunes, or at least their MPI component,
are hence specific to a particular PDF --- even, arguably, to the variation fits
within a given PDF set.

A key aspect of MPI tuning in modern MC generators is to fit the \sqrtS
dependence of the model. The \jimmy MPI model, and early versions of \herwigpp,
attempted to describe all MPI activity with a fixed \pT regulariser value at all
energies, but this ultimately proved unworkable and a \pythia-like \sqrtS
dependence with a slow power-law dependence similar to total $pp$ cross-section
fits~\cite{Donnachie:1998gm} was instead introduced. The analogy is not
predictive, however, so this exponent is also a free tuning parameter for any
tune interested in describing more than one collider energy. Experimental data
from more than one energy is obviously needed to constrain this parameter, but
the power-law ansatz has been found to work well and even been supported
empirically by independent tunes of $\pT^0$ at different
\sqrtS~~\cite{Schulz:2011qy} as shown in \fref{fig:tuning:pt0sqrts}.

Finally, a marginal aspect of tuning: the ``primordial \kT''. This quantity,
implemented in several parton shower generators, is the $\mathcal{O}(1)~\GeV$
width of a function used to add a randomly sampled transverse momentum boost to
the whole modelled scattering event. This is motivated almost entirely by the
pragmatic desire for a good description of the $Z$ \pT differential
cross-section, a precisely measured observable generated by initial-state
recoils, which rises steeply from zero at $\pT(Z) = 0$ to a peak at a few
\GeV. The exact position of this peak is determined by resummation of large QCD
logarithms, i.e.~the process approximated by the parton shower. Comparisons to
data with a ``vanilla'' parton shower almost invariably produce a peak at too
small a \pT value, and hence primordial \kT smearing was introduced as the
simplest possible mechanism to ``correct'' this flaw in data
description\footnote{While often justified via an uncertainty-principle
  ``particle in a box'' argument, the typical magnitude of the smearing width is
  an order of magnitude larger than expected from such an argument.}. While this
nebulous shortcoming of parton shower models is somewhat distressing, at present
only the $Z$ \pT distribution is precisely enough measured at hadron colliders
to be sensitive to this effect (and other very soft QCD modelling effects, such
as the ISR shower cutoff scale and/or \powheg real emission
cutoff~\cite{Frixione:2007vw}). Primordial \kT can hence be tuned virtually
independently of other initial-state quantities, although it can equally be
included in larger tunes, where it becomes a flat direction in the parameter
space for all bins except those in the low-\pT region of $Z$ \pT and related
observables.





\subsection{Parameterisation-based tunes:}
\label{sec:tuning:fitting}


Tuning via parametrisation of MC generator behaviour has a lengthy
history~\cite{Althoff:1984rf,Braunschweig:1988qm,Buskulic:1992hq,Barate:1996fi,Hamacher:1995df,Abreu:1996na}. The
fundamental idea is to replace the expensive, probably multi-day, explicit
evaluation of a proposed MC parameter point with a very fast, analytic
approximation.

It is tempting to try to parametrise the shape of an observable as a whole as
function of some input parameters by using splines or similar structures. This,
however, is a non-trivial task as the functional form of an observable will in
most cases not be parametrisable by simple functions. Similarly, parametrising
the entirety of a multi-bin goodness-of-fit (\gof) function proves fraught. Instead, a
$P$-dimensional polynomial is independently fitted to the generator response,
$\text{MC}_b(\vec{p}=(p_1,\dots ,p_P))$, of each observable bin $b$. By doing so
the potentially complicated behaviour of observables is captured by a collection
of simple analytical functions.

Having determined, via means yet undetailed, a good parametrisation of the
generator response to the steering parameters for each observable bin, it
remains to construct a \gof function and minimise it. The result is a predicted
parameter vector, $\vec{p}_\text{tune}$, which should (modulo checks of the
technique's robustness) closely resemble the best description of the tune data
that the generator can provide.

In parametrisation-based tuning, the run-time is dominated by the time taken to
run the generator to produce inputs to the parameterisation.  This step is
trivially parallelisable and large tunes can be tractable even with modest
computing resources.  The calculation of the parameterisation rarely exceeds a
few minutes, as does the subsequent numerical minimisation step: this technique
hence enables rapid tuning in response to new measurements, as well as
systematic exploration of freedoms in the tuning procedure itself. The
preparation of input data depends on the complexity of the task at hand, namely
the sophistication of the generator and the dimension of the parameter space.



\begin{figure}[t]
  \centering
  \includegraphics[width=0.45\textwidth]{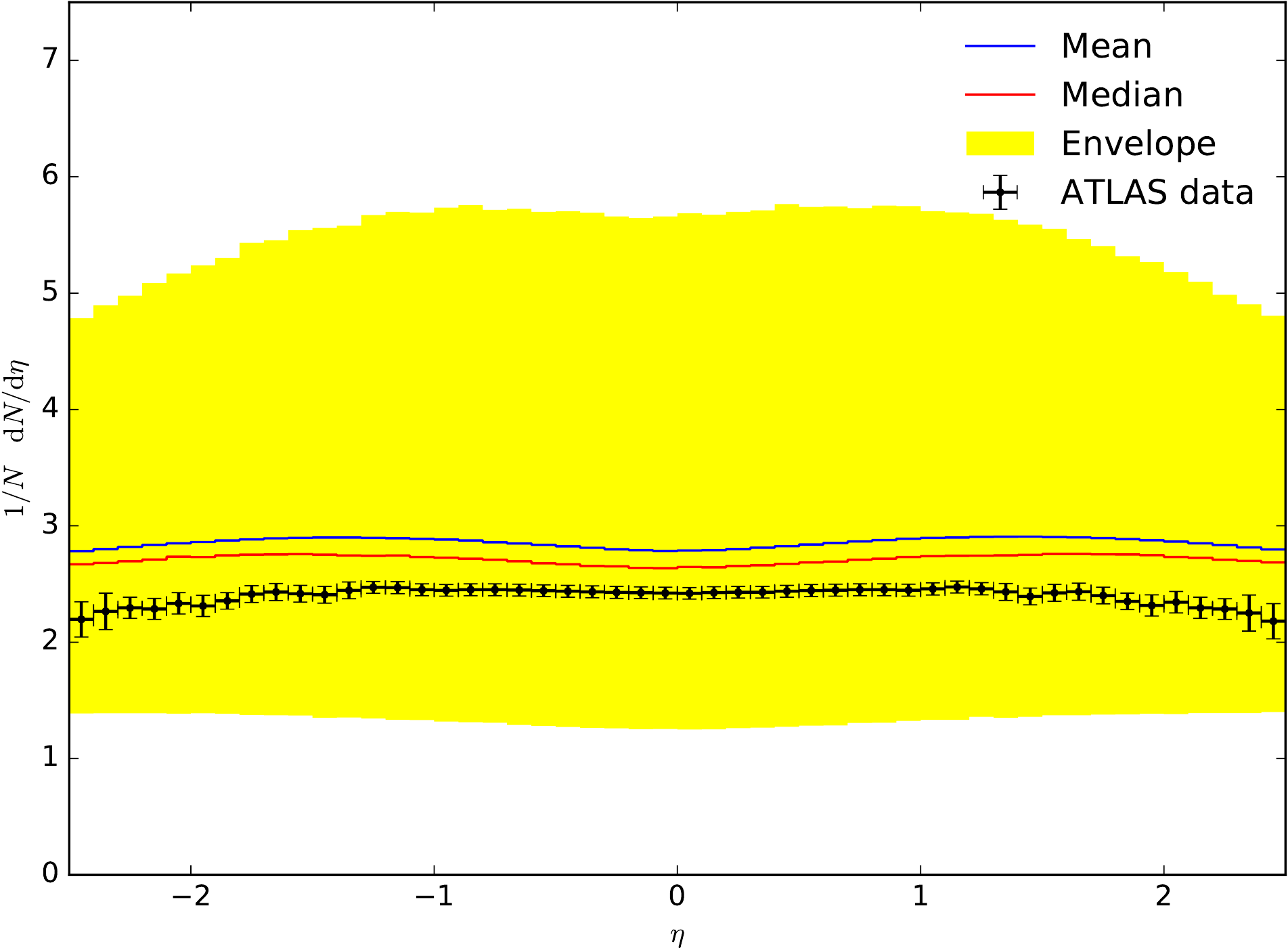}
  \caption{Example of a tuning ``envelope plot''. Here the experimental data is
    clearly contained within the yellow band of the sampled model space, but
    that is not always the case.}
  \label{fig:tuning:envelope}
\end{figure}

It is important to check the fidelity of the parametrisation, and to ensure that
the parametrisation scan includes predictions surrounding the target data
values. The construction of ``envelope plots'' is a good practice to check
\textit{a priori} that the chosen sampling range actually covers the data
considered for tuning. For each bin of each observable the minimal and maximal
value from the corresponding inputs is obtained and thus allows to spot mistakes
and model limitations early on, as shown in \fref{fig:tuning:envelope}. Using
disjoint sets of input MC data for parametrisation-building and testing allows
the accuracy of the parametrisation to be estimated.

\paragraph{Fitting model}

To illustrate the method, we discuss the parameterisation of the bin content
$\text{MC}_b$ using a general polynomial of second order:
\begin{equation}
  \text{MC}_b(\p)
  \approx  \alpha^{(b)}_0 + \sum_i \beta^{(b)}_i \, p^\prime_i
  + \sum_{i \le j} \gamma^{(b)}_{ij} \, p^\prime_i \, p^\prime_j.
\end{equation}
The task at hand is to determine the coefficients $\alpha, \beta_i, \gamma_i$.
Algorithmically this is done by generating at least as many inputs
$\text{MC}_b(\p)$ for different $\p$ as there are coefficients in the to be fit
polynomial such that we are able to solve a system of linear equations.  Since
it is much more practical we cast the right hand side into a scalar product:
\begin{equation}
\text{MC}_b(\p) \approx f^{(b)}(\p) = \sum_{i=1}^{N_\text{min}(P)} c^{(b)}_i \, \ptilde_i .
\end{equation}

\noindent For a second order polynomial of a two dimensional parameter space
($x,y$), the coefficient vector would have the form
\begin{equation}
\vec{c^{(b)}} = (\alpha, \beta_x,\beta_y, \gamma_{xx}, \gamma_{xy}, \gamma_{yy}),
\end{equation}
and for each set of input points ($x_i, y_i$) we can write
\begin{equation}
\ptilde_i = (1, x_i, y_i, x_i^2, x_iy_i,y_i^2).
\end{equation}

\noindent By doing so and denoting the set of $\ptilde_i$ as $\tilde{P}$ and the
set of corresponding bin values as $\vec{\text{MC}_b}$ we construct the matrix
equation
\begin{equation}
\text{\vec{MC}}_b =\mathbf{\tilde{P}}\cdot \vec{c^{(b)}},
\end{equation}
which allows us to determine the set of
coefficients $\vec{c^{(b)}}$ by inverting $\mathbf{\tilde{P}}$, i.e.
\begin{equation}
\vec{c^{(b)}}=\mathcal{I}[\tilde{P}]\vec{\text{MC}_b}.
\end{equation}

In \professor, a singular value decomposition (SVD) algorithm (implemented in
the Eigen3 library) is used to perform the matrix (pseudo)inversion
$\mathcal{I}[\tilde{P}]$. The SVD method is equivalent to a desirable
least-squares fit of the target polynomial to the input data. In the case of
having as many input points as there are coefficients to be determined, the
solution is exact. When providing more than $N_n(P)$ inputs, the system is
over-constrained and therefore the fit will average out to some degree both
statistical fluctuations and the fact that the true generator response will
almost never be fully describable by a general polynomial. We prefer to
oversample by at least a factor of two for robustness.  As only the central bin
values enter the SVD algorithm, but the statistical uncertainty of the input
data is crucial to \gof construction and optimisation, the bin
\emph{uncertainties} are fitted as a separate polynomial
in exactly the same way as the bin values.

\begin{figure}[t]
  \centering
  \includegraphics[width=0.45\textwidth]{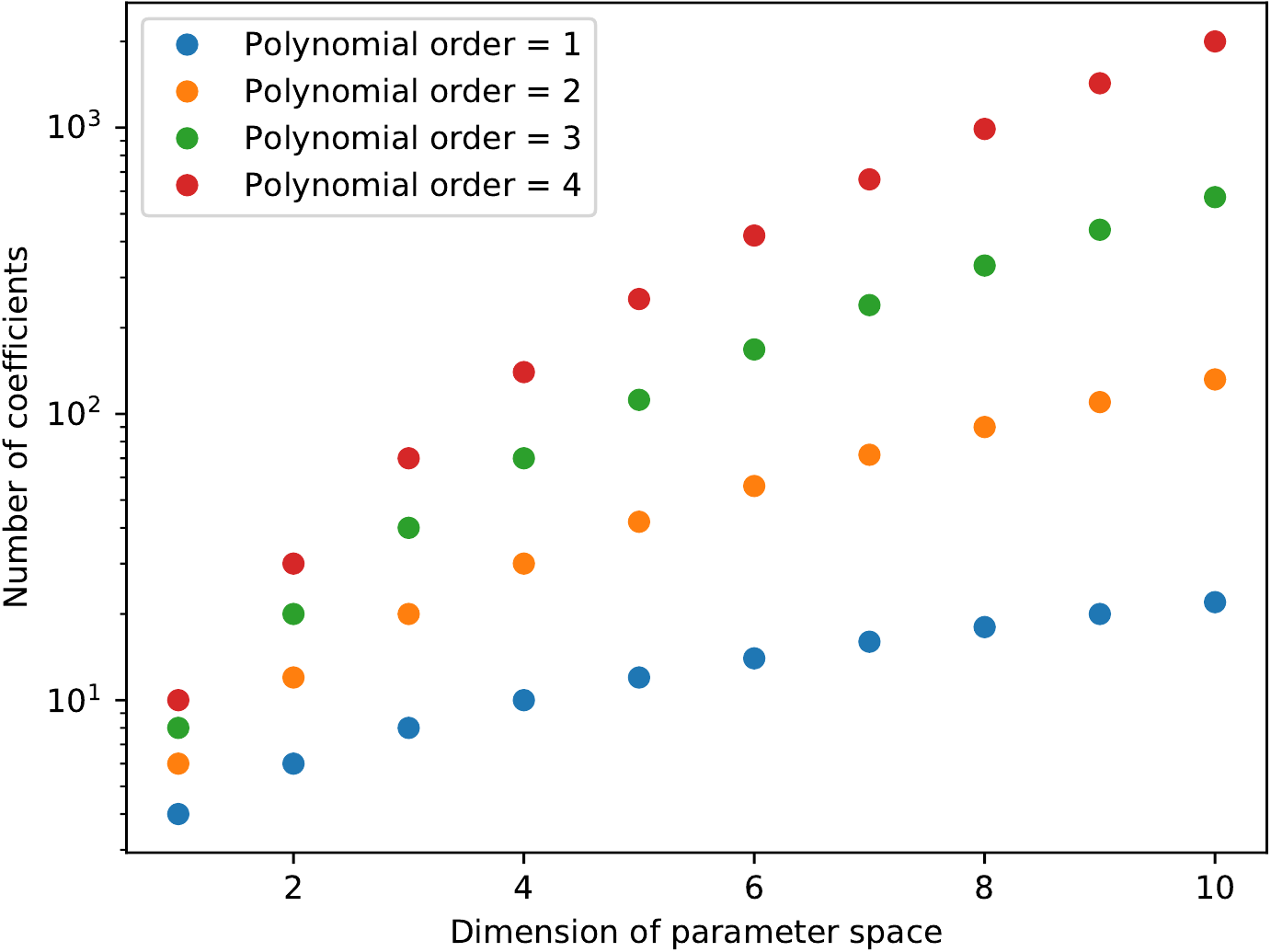}
  \includegraphics[width=0.45\textwidth]{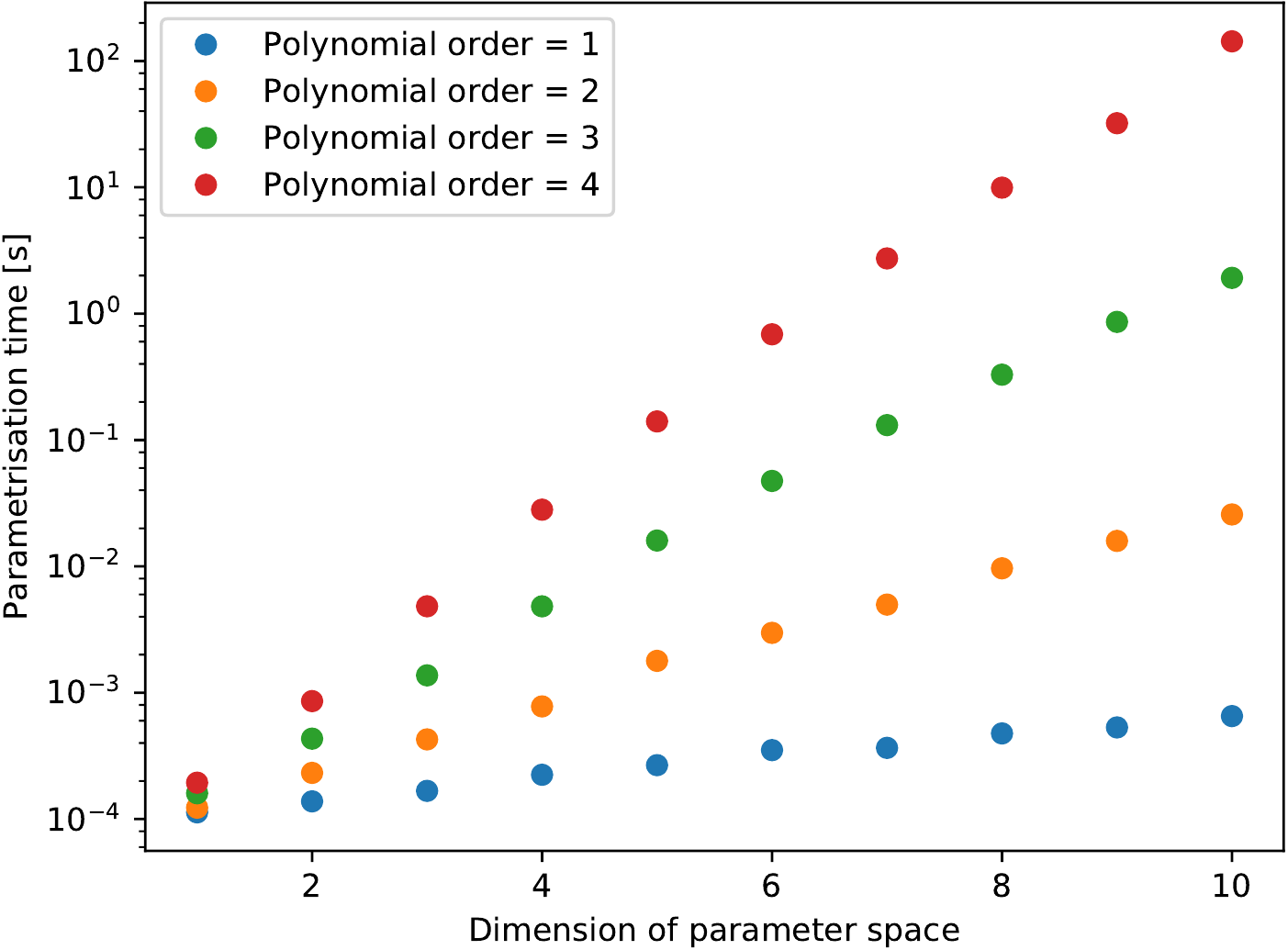}
  \caption{Parameterisation requirements. Left, $N_n(P)$ vs $P$; Right, single core computation time vs $P$.}
  \label{fig:tuning:coefftimevsdim}
\end{figure}

The dimension of the parameter space and the order of the polynomial to be
fitted determine the dimension of the to be inverted matrix and thus the
minimal number of required input datasets. Generating more inputs than
minimally necessary is in this context equivalent to over-constraining a system
of linear equations. Doing so has the benefit of being able to test the
stability of the obtained best parameter point against two aspects. One being
the order of polynomials chosen as higher order correlations can become
important. Secondly, although  parametrisations obtained from all available
inputs should give the best prediction of the generator response in the whole
of the parameter space, smaller subsets can yield different best parameter
points which is indicative of the polynomial approximation breaking down
(typically in a too large parameter space). The number of coefficients of a
$P$-dimensional general polynomial of order $n$ is
\begin{align}
  N_n(P) = 1 + \sum_{i=1}^{n} \, \frac{1}{i\,!} \, \prod_{j=0}^{i-1} (P+j).
\end{align}
How the number of parameters scales with $P$ for 2nd and 3rd order polynomials
is tabulated in \tref{tab:ncoeffs} and shown in \fref{fig:tuning:coefftimevsdim}.
The latter also shows how computational cost scales with $P$ and $n$.

\begin{table}[t]
  \centering
  \begin{tabular}[t]{lll}
    \toprule
    Num params, $P$ & $N_2(P)$ (2nd order) & $N_3(P)$ (3rd order) \\
    \midrule
    1   & 3         & 4   \\
    2   & 6         & 10  \\
    4   & 15        & 35  \\
    8   & 45        & 165 \\
    \bottomrule
  \end{tabular}
  \caption{Scaling of number of polynomial coefficients $N_n(P)$ with dimensionality
    (number of parameters) $P$, for polynomials of second- ($n=2$) and third-order ($n=3$).}
  \label{tab:ncoeffs}
\end{table}

\professor allows to calculate polynomials of in principle arbitrary order,
including $0$th order, i.e. the constant mean value.  To account for
lowest-order parameter correlations, a polynomial of at least second-order
should be used as the basis for bin parametrisation.  In practice, a 3rd order
polynomial suffices for almost every MC generator distribution studied to date,
i.e. there is no correlated failure of the fitted description across a majority
of bins in the vicinity of best generator behaviour.  An upper limit on the
usable polynomial order is implicitly given by the double precision of the
machinery effectively limiting the maximum usable order to about 11. It should
be noted that other classes of single polynomials such as Chebychev or Legendre
have no additional benefit in the \professor method. The quality of the
resulting parametrisation is exactly the same, with merely different groupings
of the coefficients.

The set of input points for each bin are determined by randomly sampling the
generator from $N$ parameter space points in a $P$-dimensional parameter
hypercube $[\,\p_{\text{min}}, \p_{\text{max}}]$ defined by the user. This
definition requires physics input --- each parameter $p_i$ should have its upper
and lower sampling limits $p_{\text{min,max}}$ chosen so as to encompass all
reasonable values while avoiding discontinuities.  In cases where bin values
vary over many orders of magnitude general polynomials are obviously a poor
choice for an approximate function. It is nonetheless possible to use \professor
by simply taming the input data by parametrising e.g. the logarithm of bin values.

\paragraph{Goodness of fit function and optimisation}


With the parameterisation of the generator at hand the optimisation can be
turned into a numerical task.  By default an \emph{ad hoc} \gof measure is
minimised using Minuit's migrad algorithm with the typical features of being
able to fix or impose limits on individual parameters during the fit.  For the
purpose of generator tuning with typically imperfect models it is often
necessary to bias the \gof in order to force the description of certain key
observables, e.g. the plateau of the underlying event or to switch off parts of
histograms entirely when it is known that the underlying model is unable to
describe the data at all.  The relative importance of individual bins and
observables can be set in \professor using weights, $w_b$, in the \gof
definition:
\begin{align}
  \chisq(\p) =
  \sum_{\mathcal{O}}  \sum_{b \, \in \, \mathcal{O}} w_b \cdot
  \frac{ (f^{(b)}(\p) - \mathcal{R}_b)^2 }{ \Delta^2_b (\p)},
  \label{eq:chi2}
\end{align}
where $\mathcal{R}_b$ is the reference data value for bin $b$. The error
$\Delta_b$ is composed of the total uncertainty of the reference which is a
constant for each point \p and the parameterised statistical uncertainty of the
Monte Carlo input for bin $b$.  In practice we attempt to generate sufficient
events at each sampled parameter point that the statistical MC error is much
smaller than the reference error for all bins.

\paragraph{Further methods}

Optimisation in tuning is by no means limited to this setup. The Python language
bindings to the core objects in Professor allow to easily construct arbitrary
\gof measures and to use other minimisers that have Python bindings, such as
MultiNest~\cite{Feroz:2013hea}.

Although being highly successful, the polynomial parameterisations have obvious
limitations. The framework hence allows to use different parameterisation
methods for instance the ones available in \texttt{scikit-learn} and
\texttt{Gaussian Processes} where no prior assumption on the functional form has
to be made.

\professor also provides tools for the interactive exploration of the parameter
space as GTK application as well as Jupyter notebooks. Both allow to load a
previously calculated set of parametrisations and displays the corresponding
histograms. GUI sliders (one per parameter) allow to conveniently set the parameter
point to a new value resulting in the histogram being redrawn immediately. These
tools help to build intuition into what effect parameters have on distributions.

\section{Influential MC generator tuning results}
\label{sec:tuning:tunehistory}

Parton shower MC generators have been tuned for as long as they have existed:
the intrinsic limitations of their formal accuracy and the dependence of
truncated perturbative calculations on unphysical scales means that some
authorial intuition has long been needed to achieve a good description of key
data observables ``out of the box''. At LEP, the experiments also sank effort
into tuning of final-state parton showers and
fragmentation. 
But the existence of a wider programme of tuning, particularly for MPI
modelling, started with the work of Rick Field and the CDF tunes of
\pythiasix~\cite{Field:2006gq}. 

The first such construction was Tune~A, its name acknowledging immediately that
it was bound not to be the last word on this limitless subject. Tune~A was
constructed specifically to provide a first reasonable modelling of the
underlying event, based on CDF's first UE
measurement~\cite{Affolder:2001xt}. However, it was soon noticed that its
parameter choices, particularly for the \mbox{PARP(64)} parameter governing the ISR
evolution scale and the \mbox{PARP(91)} primordial \kT width, resulted in a $Z$ \pT
spectrum whose peak was at too small a \pT value. A new tune was
needed~\cite{Field:2006gq,Affolder:1999jh}, and soon came in the form of Tune~AW
with a 2.1~\GeV primordial \kT width and a reduced ISR scale --- corresponding
to a larger \alphaS and hence more initial-state radiation against which to
recoil. Inevitably, Tune~AW also hit an obstacle: the dijet azimuthal
decorrelation distribution~\cite{Abazov:2004hm}, characterising the extent to
which, in generic Tevatron jet events, initial-state recoils between high-\pT
jets push the leading dijet system away from the 2-body back-to-back
configuration. Having specifically boosted the amount of ISR to create Tune~AW,
it now needed to be reduced again (this time via \pythia's \mbox{PARP(67)} ISR
starting scale parameter) to avoid creating too many hard multi-jet events: the
resulting tune was christened Tune~DW. Several other variations joined the
swelling ranks of CDF \pythiasix tunes, including variations in PDF choice and
the energy dependence of MPI regularisation. (This last issue, almost in
isolation, was also applied to tuning of the \herwig MPI model, \jimmy, in
anticipation of the energy leap from Tevatron to LHC.)

The pattern at this point had become clear: the initial-state system of MPI,
ISR, and intrinsic \kT --- as well as developments in matrix element matching \&
merging --- was too complex to be entirely optimised by hand. Every
single-parameter change would both address the modelling problem at which it was
aimed, and break several other distributions. The \professor tuning effort arose
at this point, to apply the computational methods described above to this
optimisation problem. The first and only tunes to bear the \professor label were
the Prof0-Q2 and Prof0-pT tunes~\cite{Buckley:2009bj} of \pythiasix, for its
virtuality ordered parton shower and newer \pT-ordered parton shower
respectively. Both tunings were ``global'', in the senses that they covered all
aspects of the generator from final-state showering and hadronisation, to the
initial-state effects covered by the CDF tunes\footnote{The Prof0-pT tune in
  fact provided the first final-state tune of the \pythiasix \pT-ordered shower,
  which had been previously used with the virtuality-ordered settings.} as well
as the widest available dataset from LEP hadron spectra to event shapes, and to
Tevatron minimum-bias and underlying event data. More influentially, the
\professor machinery was immediately used within ATLAS to produce its own tune
series first based on CDF data and then including the early ATLAS data in the
AMBT and AUET tune series between 2009 and 2012. 
Also influential were the hand-tuned ``Perugia'' family by Peter~Skands,
particularly since they included systematic variations of the parton showers
useful for uncertainty estimation.

\begin{figure}[tp]
  \centering
  \includegraphics[width=0.45\textwidth]{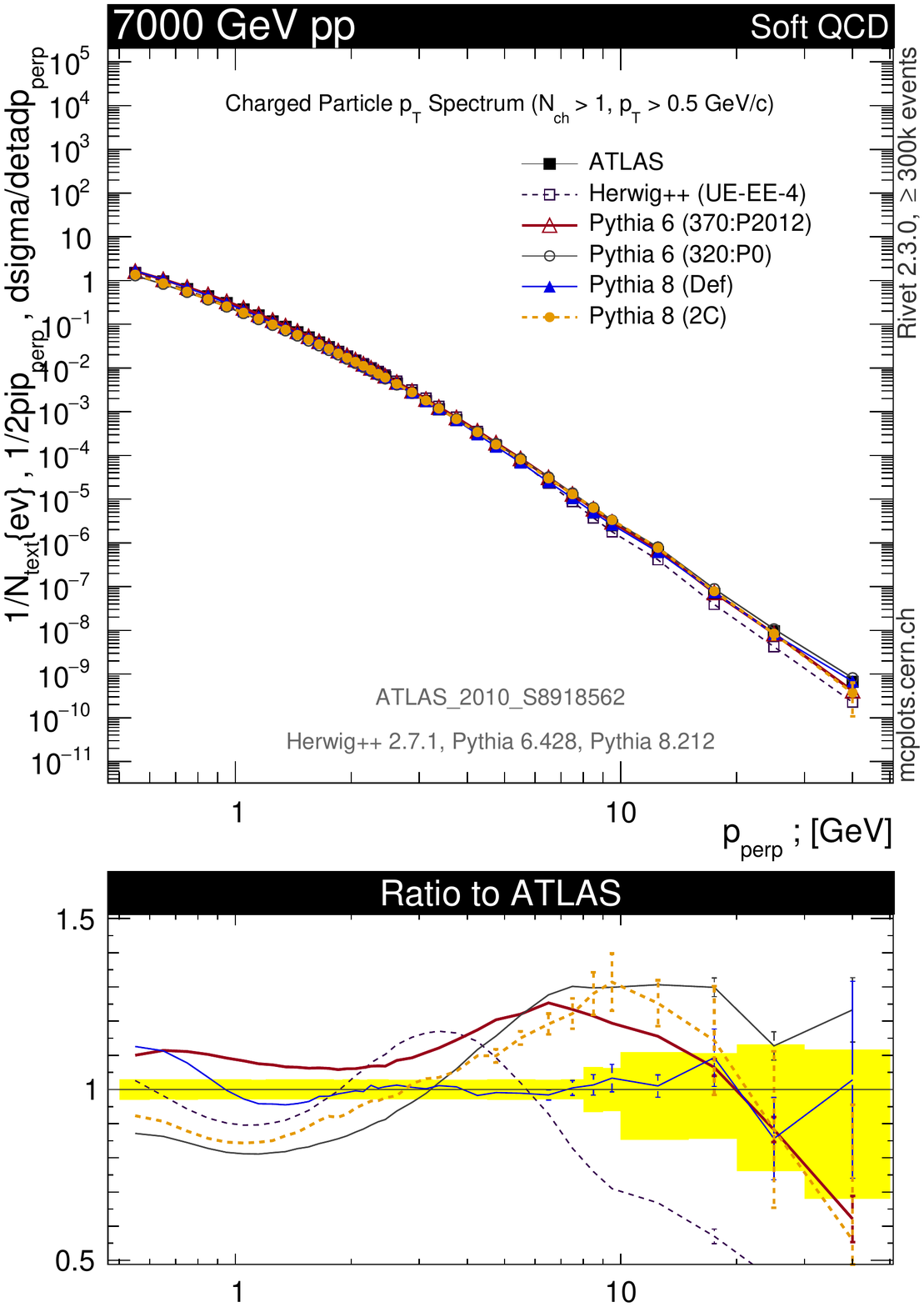}
  \includegraphics[width=0.45\textwidth]{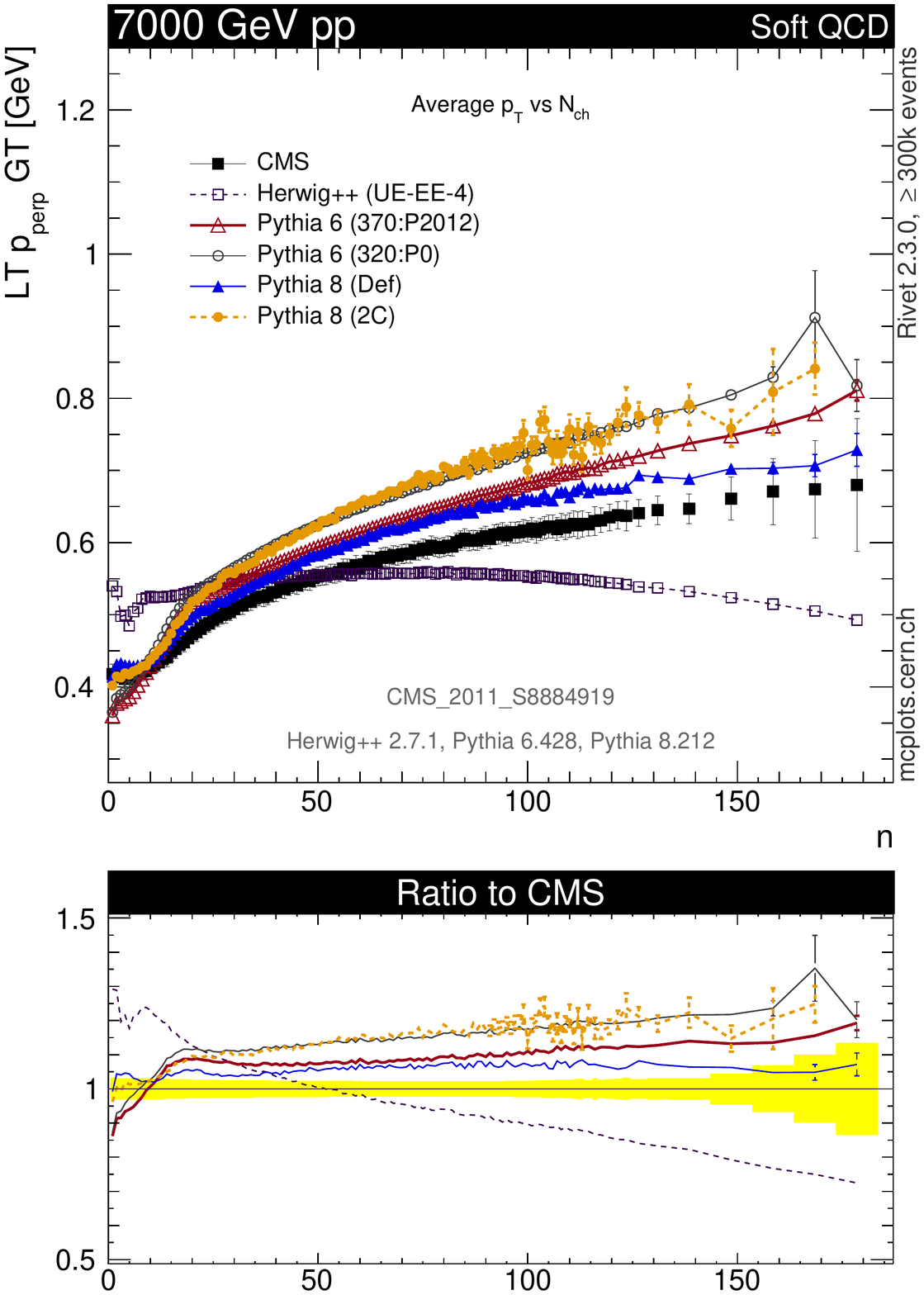}\\[1em]
  \includegraphics[width=0.45\textwidth]{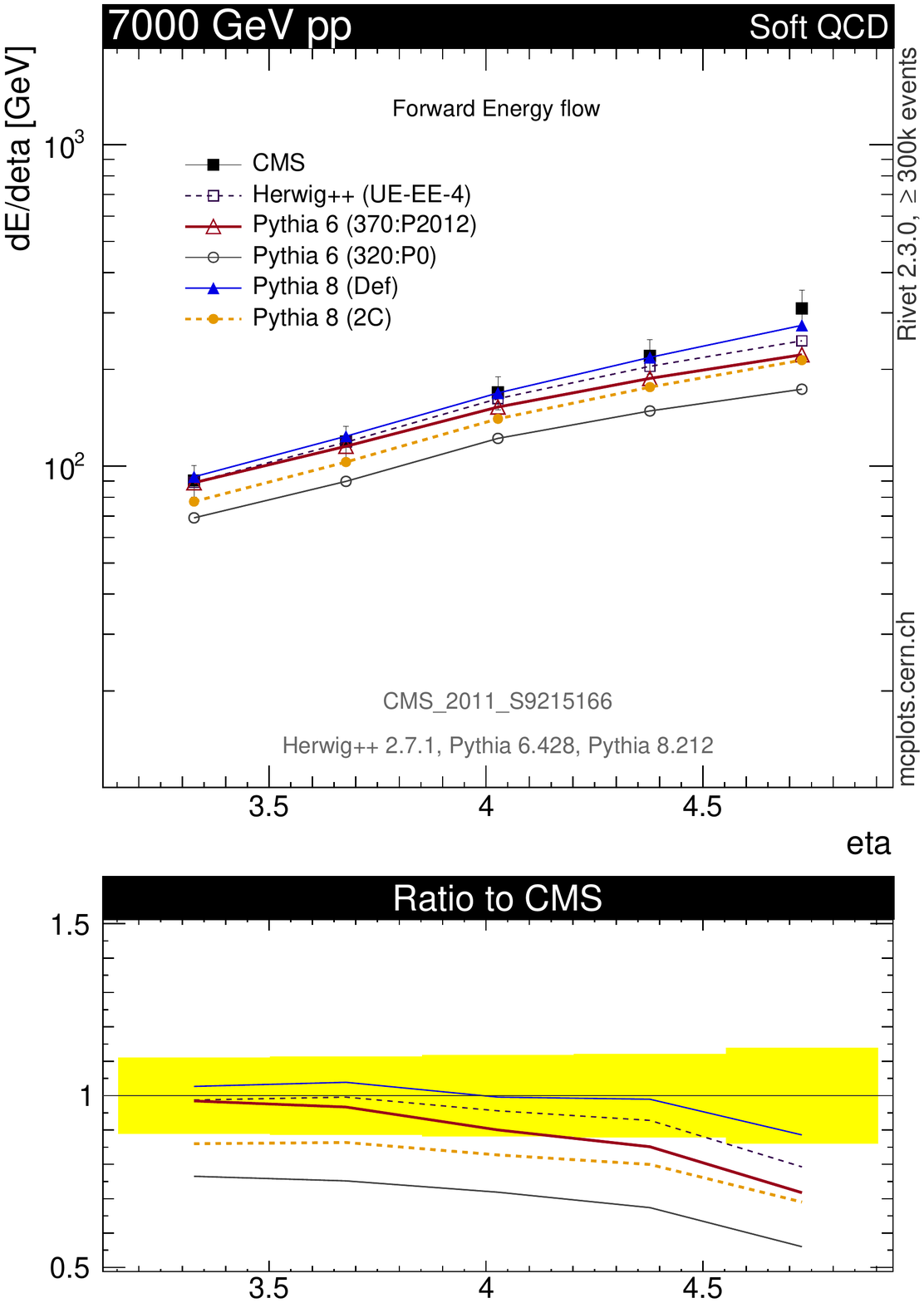}
  \includegraphics[width=0.45\textwidth]{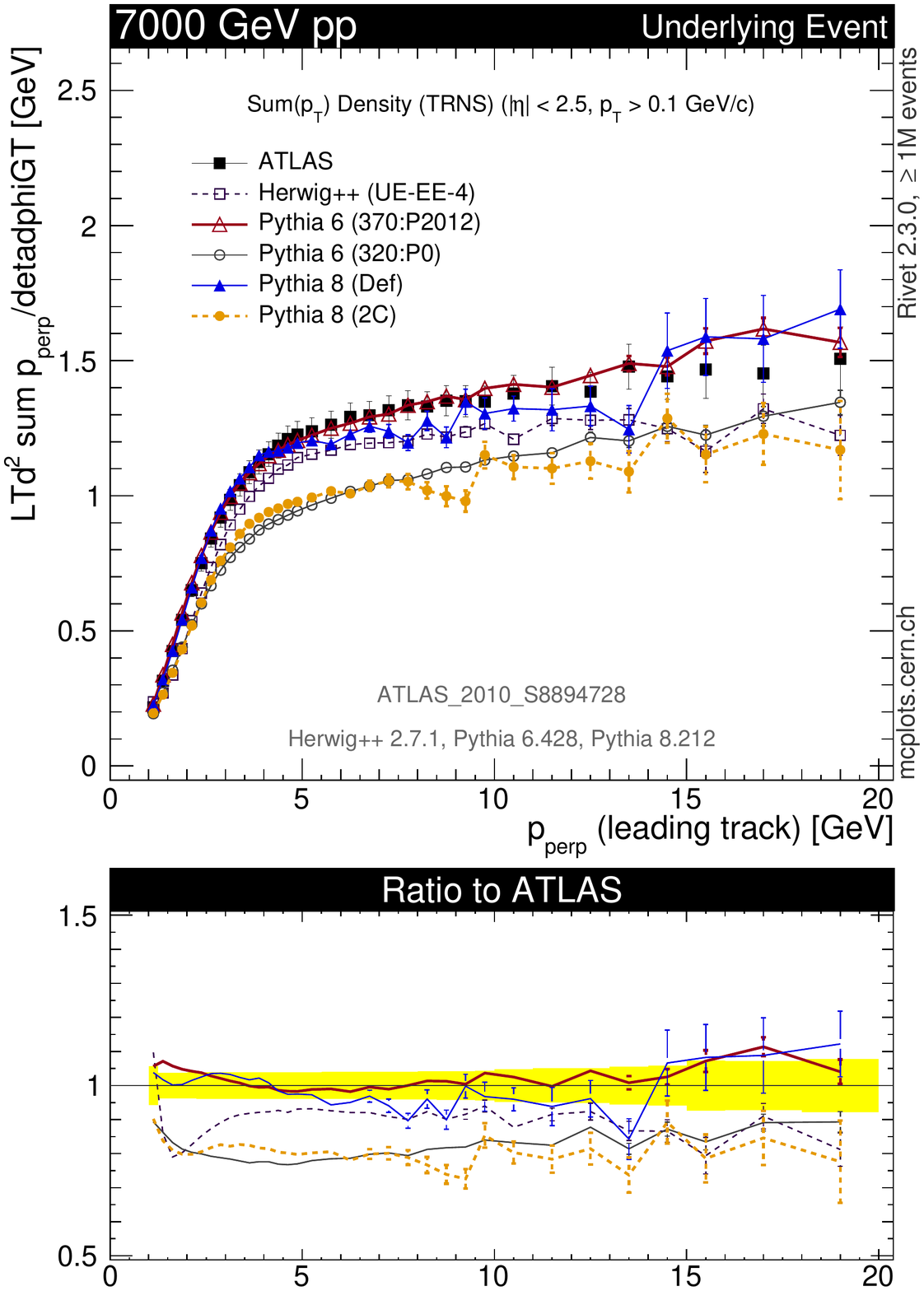}
  \caption{Performance of MC tunes on MPI-sensitive observables, showing
    \pythiasix (two ``Perugia'' tunes), \pythiaeight (the author Tune~2C and
    default Monash tune), and \herwigpp (UEEE-4 tune). The Perugia\,0 and 2C
    tunes are based on pre-LHC \tevatron data, and the others include early LHC
    data. The top row shows LHC minimum-bias \pT spectrum and
    $\langle \pT \rangle$ vs.~$N_\mathrm{ch}$ observables, and the bottom row
    shows tune performance on minimum-bias energy flow (left), and underlying
    event $\sum \pT$ observables. The inclusion of LHC data in all cases
    improves the data description.}
\end{figure}

\afterpage{\clearpage}

As suggested by the story so far, the initial LHC tuning community focus was
concentrated on \pythiasix. Additional work was performed at this time, largely
via the \professor technique, to tune the \sherpa fragmentation model and the
\jimmy MPI mechanism for the \herwigsix generator. The next major developments
in MPI tuning were the shift during LHC Run~1 to the newer C++ family of
generators.  The \pythiaeight, \herwigpp, and \sherpa generators were all tuned
using the \professor tools within their development collaborations, with the
most notable outputs being the \pythiaeight Tune~4C, and \herwigpp's UEEE tune
series which introduced tuned modelling of MPI energy evolution and
colour-reconnection in response to the evidence that MPI observables could not
be successfully tuned within the existing model space without such
mechanisms.

ATLAS and CMS tuning of \pythia continued, with ATLAS's A1, A2 and AU2 tunes
being heavily used in the Run~1 simulation leading up to the Higgs boson
discovery, while CMS' ``Z''-tune variants on the AMBT series were used for the
same purpose on that experiment. Each experiment focused on its own growing
collection of soft-QCD data analyses as the LHC energy increased. At the end of
LHC Run~1, Skands and collaborators provided the ``Monash'' global tunes of
\pythiaeight~\cite{Skands:2014pea}, which ATLAS modified into the ``A14'' tunes
for use in Run~2 modelling, incorporating high-\pT $Z$ and $t\bar{t}$
observables into the fit to serve the needs of BSM
searches~\cite{ATL-PHYS-PUB-2014-021}. CMS, meanwhile, constructed its own Run~2
series, the CUET and CDPST tunes~\cite{Khachatryan:2015pea}, the latter of which
is unique in being tuned to hard double-partonic scattering data. At the time of
writing, these experiments' and authors' tunes of \herwigpp and \sherpa are the
most widespread general purpose tunes at the LHC.

Recently most LHC tuning effort has been focused on configurations most suitable
for use with matching and merging event generators in which the parton shower is
interleaved with collections of higher-order matrix elements. The results have
been specialist tunes such as ATLAS' AZ and AZNLO~\cite{ATL-PHYS-PUB-2013-017}
(specifically for description of the $Z$ \pT, to be used in $W$-mass
measurement) and ATTBAR~\cite{ATL-PHYS-PUB-2015-007}. The LHC split between
tunes for minimum-bias data description and underlying-event description has
also still to be resolved, perhaps by inclusion of more advanced diffractive
physics models although efforts along those lines have yet to prove fully
satisfying.








\section{Tuning uncertainties}
\label{sec:tuning:uncertainties}

\begin{figure}[tp]
  \centering
  \subfigure[][]{\label{fig:tuning:ueextrapolation}\raisebox{0em}{\includegraphics[width=0.55\textwidth]{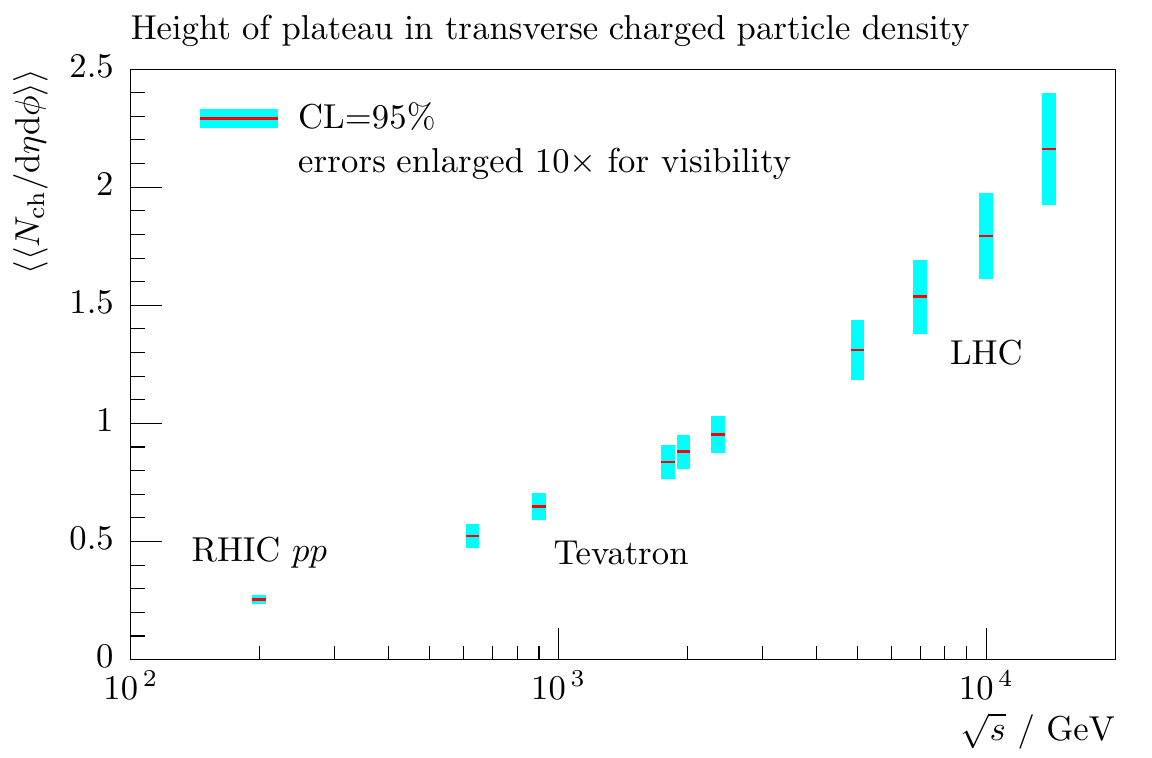}}}
  \subfigure[][]{\label{fig:tuning:a14eigentunes}\includegraphics[width=0.44\textwidth]{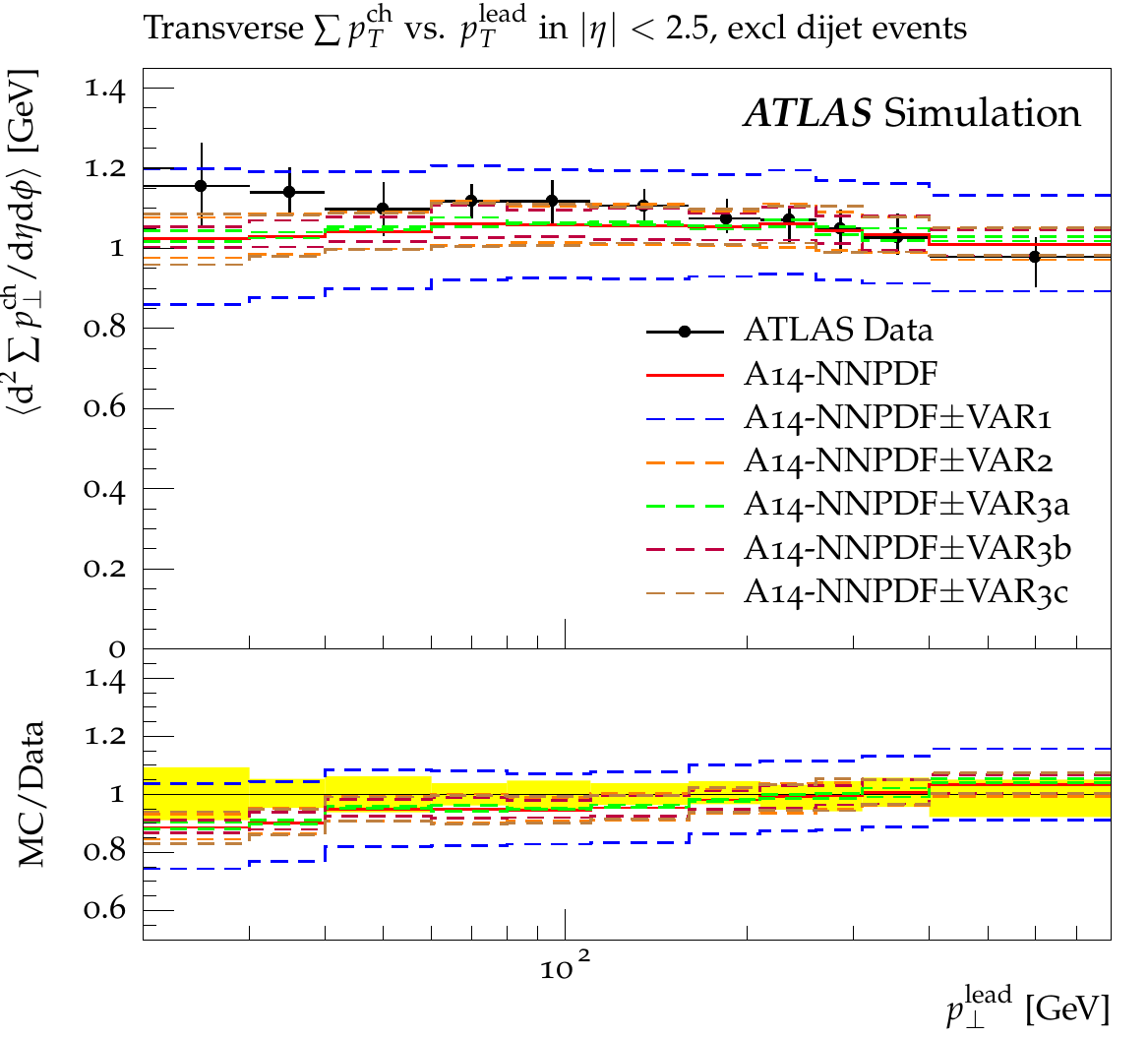}}
  \caption{Use of tune uncertainty estimation to predict LHC underlying event activity}
\end{figure}

Producing optimal fits of MC models to data is valuable, but not the whole
story. In particular for experimentalists' usage of these simulations, it is
crucial that the \emph{uncertainties} in a tuned model also be quantified. This
permits, in varying degrees of sophistication, numerical treatment of modelling
uncertainties as nuisance parameters in fits of physics both from the Standard
Model and from beyond it. Estimates of tuning uncertainty also proved valuable
in the run-up to LHC operations, when tune fits to Tevatron and other low-energy
data permitted a quantitative estimate of the range of underlying event activity
to be expected at the new 13~\TeV collider --- this extrapolation of uncertainty
is visible in \fref{fig:tuning:ueextrapolation}.

Simple uncertainty estimates can be produced in various \emph{ad hoc}
ways. First, an approach analogous to scale choices in perturbative calculation:
simply pick some parameter variations that seem ``reasonable'' --- typically
factors of two --- and release them as variation tunes. A step up in
sophistication is to make such changes to key parameters, e.g. enforcing more or
less initial-state radiation, and then re-tuning the remaining $(P-1)$-parameter
system to infer how much other parameters can compensate for the forced move
away from the global optimum. As with scale-setting, there is a degree of
artistry to this: parameter changes which result in very large or very small
changes to observables may be judged as ``unreasonable'' and be modified
accordingly.

The logical conclusion of this is to decide that the ultimate arbiter of how
large a parameter change should be is that it produces a model variation
comparable to the measured experimental uncertainty on the observables, such
that the union of all parameter variations envelopes all or at least most data
uncertainties.  This approach is the philosophy adopted by several recent
\professor-based tunes, such as the A14 series~\cite{ATL-PHYS-PUB-2014-021},
which provide ``eigentune'' variations to complement the global fit --- an
example is shown in \fref{fig:tuning:a14eigentunes}.

The additional detail in eigentune construction is that there are an infinity of
ways in which to make model variations ``cover'' all data, so we prefer a set of
variations which are maximally decoupled from one another. This can be achieved
by use of a second-order approximation to the $\chi^2$ valley around the global
optimum: in general this will be an ellipsoid in the parameter space, and the
principal basis of this ellipsoid defines $2P$ principal vectors along which to
make parameter variations. These vectors are obtained as the eigenbasis of the
covariance matrix computed at the global tune point.

If the test statistic were truly distributed according to the $\chi^2$
distribution, the required deviation could be calculated analytically from the
$\chi^2(k,x)$ distribution for degree of freedom $k = N_\mathrm{bins}-P$, by
making pairs of deviations along the $P$ eigenvectors until the $\chi^2$
corresponding to a $p$-value of e.g. $1\sigma$ is found: for a perfect $\chi^2$
statistic and a best-fit value of $\chi^2_\mathrm{best} \sim k$, this should
require a $\Delta\chi^2$ shift of $\sim \! 2k/3$. But in practice we find that
such a construction fails the ``reasonableness test'': the resulting variations
are far too small. Empirically, the typical distribution in global LHC fits has
a much larger mean than expected for the number of degrees of freedom, and a
narrow spread incompatible with the $\chi^2$ distribution's relation between
mean and variance. Instead the strategy of producing empirical model variations
which cover the experimental uncertainties has been adopted in tune sets such as
A14, requiring a $\Delta\chi^2 \sim N/2$ to produce experimentally useful
systematic variations. Other, less empirical, approaches are still being explored.

It is thought that this deviation from statistical expectation largely stems
from the fact that we do not yet have models capable of reproducing all data
observables: ``the truth'' is not contained in the model space. In addition, the
data available so far for MPI tuning have not included detailed bin-to-bin
correlations from systematic uncertainties, which in principle could be
eliminated by a nuisance parameter fit --- this futher distorts the
goodness-of-fit measure away from the $\chi^2$ distribution. Finally, there is
the ever-present risk of underestimated experimental uncertainties. There is
hence still potential to improve tuning methodology, by construction of more
robustly motivated systematic variations and reducing tuning uncertainties by
fully
correlated 
likelihood construction across multiple measurements.

In addition to the eigentune method described here, a full assessment of tuning
uncertainties should coherently encompass statistical uncertainties in the
\professor parameterisation construction (estimated by making many
semi-independent parameterisations, from subsets of the available MC runs), and
correlations in $\chi^2$ construction. Accurate treatment of these effects may
eventually permit tune uncertainties to be treated on the same statistical level
as those of parton density function fits.

\section{Outlook}
\label{sec:tuning:outlook}

Tuning has proven to be an important activity in the development of MC models of
initial-state QCD at the LHC, both by providing the experiments with
unprecedentedly well-honed simulations of collision events (including pile-up),
but also by providing a mechanism by which to unambiguously identify when a
model's limitations are fundamental. At the same time, the development of tuning
machinery for the LHC has provided ways to quantitatively estimate model and
tune uncertainties, and in principle to reduce them --- although the statistical
foundation still requires development.

Technically, the \professor framework has seen recent advances, originally
developed for BSM physics studies but readily applied to QCD MC tuning: the most
obvious of these are the inclusion of non-polynomial functional forms such as
neural nets, suport vector machines, and methods based on decision trees. Work
on using Gaussian processes, along with more refined statistical testing of
parameterisation fidelity, offer the possibility of yet more accurate MC
parameterisation for use in fitting. In parallel, a serial Markov chain approach
to tuning, based on Bayesian parameter optimisation has been developed and
appears interesting, if yet unproven on the large-scale problem of initial-state
QCD tuning where the MC runs are very computationally expensive~\cite{Ilten:2016csi}.

The most painful price paid for the increasing LHC demands of simulation
accuracy has been the proliferation of tunes specific to process types:
underlying event vs. inclusive minimum-bias, or QCD-singlet vs. coloured
hard-processes. Such fragmentation is undesirable because it implies a lack of
predictivity in the models: if we cannot trust an MPI+shower model to
simultaneously describe minimum-bias and underlying event, how confident can we
be about its extrapolation to more rarified regions of phase-space? The
resolution of this problem, and the coherent integration of diffractive
processes and hence the connection between fiducial and total inelastic
scattering cross-sections, must be the main challenges for development and
tuning of MC models in the coming years. While technology has helped the
development of tunes through the early phase of the LHC, in the end it must be
coupled to physics insights to achieve the goal of truly comprehensive
description of hadronic initial-state interactions.











\bibliographystyle{ws-rv-van}
\bibliography{tuning-ws,tuning-ws-notes}


\end{document}